\documentstyle[12pt,fleqn]{article}
\newlength{\dinwidth}
\newlength{\dinmargin}
\newcommand{\resection}[1]{\setcounter{equation}{0}\section{#1}}

\newcommand{\appsectio}[1]{\setcounter{section}{0}
         \addtocounter{section}{1} \setcounter{equation}{0}
                         \section*{#1}}
\newcommand{\appsection}[1]{\addtocounter{section}{1} \setcounter{equation}{0}
                         \section*{#1}}
\renewcommand{\theequation}{\thesection.\arabic{equation}}
\newcommand{\f}[2]{\frac{#1}{#2}}

\begin{document}
\def\ra{\rightarrow}
\def\kt{k_\perp}
\def\dt{\mbox {\boldmath $\Delta$}}
\def\pmu{p_\mu}
\def\ds{\displaystyle}
\def\tb{\bar{\tau}}
\def\ep{\varepsilon}
\def\pbmu{\bar{p}_{\mu}}
\def\pbnu{\bar{p}_{\nu}}
\def\pb{\bar{p}}
\def\k{{\bf k}}
\def\as{\alpha_s}
\def\q{{\bf q}}
\def\FF{{\cal F}}
\def\ss{\sigma}
\def\ssh{\hat{\sigma}}
\def\gvga{g_V^2+g_A^2}
\def\cvca{C_V^2+C_A^2}
\def\bks{\!\!\!}
\def\th{\hat{t}}
\def\uh{\hat{u}}
\def\sh{\hat{s}}
\newcommand{\be}{\begin{equation}}
\newcommand{\ee}{\end{equation}}
\newcommand{\bea}{\begin{eqnarray}}
\newcommand{\eea}{\end{eqnarray}}
\newcommand{\nn}{\nonumber}

 \begin{flushright}DFF 239/11/95 \\ hep-ph/9512309\end{flushright}
 \vspace*{1.5 cm}
 \begin{center}
 {\Large \bf Small-$x$ Resummation Effects in Electroweak\\
 \vspace*{.6 cm} Processes}\\
 \vspace*{1 cm}
 {\large G. Camici and M. Ciafaloni\\ \vspace*{3 mm}
 {\em Dipartimento di Fisica, Universit\`a di Firenze\\ \vspace*{3 mm}
 and INFN, Sezione di Firenze}}
 \end{center}
 \vspace*{1.3 cm}

 \begin{abstract}
 We investigate small$-x$ resummation effects in QCD coefficient
functions for $Z_0g$ and $Wg$ fusion processes, and we compare them with the
known ones of $\gamma g$ type. We find a strong process dependence, that
we argue to be due to the possible presence of collinear singularities for
either small or large $\k$ of the exchanged gluon. For top quark production,
we find that the $gg\ra t\bar{t}$ and $Z_0g\ra t\bar{t}$
channels have larger resummation enhancements than the $Wg\ra t\bar{t}$ one.
 \end{abstract}
 \cleardoublepage

\resection{Introduction}
In the past few years, various applications have been developed
[1-5] of the $\k$-factorization method \cite{1, 2} to combine
high-energy behaviour in QCD with the renormalization group $\left(RG\right)$.

On one hand, for small$-x$ hard processes of strongly interacting partons,
resummation formulae have been derived for various coefficient functions
\cite{1}
and for the next-to-leading quark entries \cite{3} of the singlet
anomalous dimension matrix. Resummation effects turn out to enhance scaling
violations in the HERA energy range \cite{6, 7}, and could be responsible
for the small$-x$ rise of structure functions seen experimentally \cite{8}.

On the other hand, $\k$-factorization is also relevant for the spontaneously
broken electroweak theory \cite{4}, where it provides a generalization of the
effective $W$ approximation \cite{9} for high-energy weak boson fusion
processes,
like, e.g., $Z_0g \rightarrow t\bar{t}$ and $Wg \rightarrow t\bar{b}$.
In such cases, it
takes into account correctly important high-energy off-shell effects which are
neglected in the (na$\ddot{\imath}$ve) collinear approximation \cite{10}.

The purpose of the present note is to calculate in detail the coefficient
functions of $Z_0g$
 and $Wg$ fusion type by concentrating on gluon off-shell effects,
and to compare the corresponding resummation formulae with the one of
$\gamma g$ type previously found. Note that while for a virtual photon source
both transverse and longitudinal coefficient functions are available, here we
limit ourselves to longitudinal gauge bosons, whose exchange dominates the
fusion process, and is related - by the off-shell equivalence theorem
\cite{4,11} -
to the corresponding Goldstone boson exchange.

It will turn out eventually that resummation effects are strongly process
dependent, and do not enhance much the $Wg \rightarrow t\bar{b}$ channel
compared
to, say, the $Z_0 g$, $gg \rightarrow t\bar{t}$ ones for top production.
Part of the
paper is devoted to explore the reasons for such a fact, and also to provide a
quick way to estimate enhancement factors in the various cases.

Roughly speaking, off-shell effects yield a cross-section increase because the
phase space for the exchanged gluon transverse momentum $\k$ opens up at
high-energies, and is furthermore weighted by the anomalous dimension exponent
$\gamma_N \simeq (3 \alpha_s/\pi) \left( N-1\right)^{-1}$, $N$ being the
moment index. Therefore, for large enough energy $s$, the integration region
where $|\k|$ exceeds the hard scale $Q$ ($Q^2 \ll {\k}^2 \ll s$) becomes
increasingly important, if not suppressed by the squared matrix element.

Let us refer to the phase space region $Q \leq |\k|$ as "the disordered
$\k$" region,
to distinguish it from the normally ordered region $Q_0 \leq |\k| \leq Q$,
typical
of the incoming parton jet. We find that the size of the cross-section
enhancement at high energies is related to the possible existence of a
collinear singularity in the disordered $\k$ region, where the exchanged quark
aligns to the hard probe, rather than to the parton in the opposite direction.
We also give the splitting functions of GLAP type which describe quantitatively
the hard off-shell cross section, thus generalizing what already noticed in the
photoproduction case \cite{1}.

For instance, while the DIS hard cross section $\hat{\sigma}_L = \sigma \left(
\gamma_L g \rightarrow q \bar{q}\right)$ - due to longitudinal photons - is not
particularly enhanced, the one for $Z_L g \rightarrow t \bar{t}$ is instead
normally enhanced, because of the collinear behaviour of the Goldstone boson
$\rightarrow Q \bar{Q}$ process. The corresponding splitting function yields a
reasonable estimate of the enhancement factor.

On the other hand, $ \hat{\sigma} \left( W_L g \rightarrow t \bar{b} \right)$
is
not particularly enhanced, despite the existence of similar collinear
properties of the $W-$Goldstone boson. But in this case, since $ m_b \ll m_t$,
the small $\k$-region is also enhanced by a $\log \left(m_t / m_b\right)$
factor due to the customary collinear singularity in the parton jet. Thus the
ratio of large $\k$ to small $\k$ effects becomes of order unity.

The contents of the paper are as follows. In Sec. 2 we set up the calculation
of the weak boson-gluon coefficient functions on the basis of our previous
treatment of $\k$-factorization in electroweak processes. The actual analytical
computation is performed in Sec. 3, and in Sec. 4 we compare the present
results with the ones for the photon source. Here we also discuss the role of
the disordered $\k$ region, and we provide the relevant splitting functions in
the various cases. Our results are discussed in Sec. 5, and some analytical
details are left to Appendices A and B, where we also discuss the relevant
longitudinal splitting functions.

\resection{$\k$-Factorization for V-g fusion processes}

We consider here contributions to heavy quark production in which the fusion
subprocess is of the mixed type, e.g. $Z_0g\ra t\bar{t}$, $Wg\ra t\bar{b}$,
the hard scale being essentially the top quark mass $m_t$ (Fig.1). Since we
have already
discussed \cite{4} off-shell effects in the weak-boson channel for on-shell
gluons, we concentrate here on the strong interaction effects, and the
correspondig resummation formulae of QCD corrections.

The fusion process that we consider (Fig. 1) is, to start with, described by
double-$\kt$ factorization, in which the hard subprocess cross section
$\hat{\sigma} \left( V\left( q\right) g\left( k\right) \ra Q_1 \bar{Q_2}
\right)$
is factorized from the structure functions in the hadron (lepton) at hand. This
means that in the high energy regime
\be
s\gg {m_i}^2, {\q}^2, {\k}^2, ~~~~~
\frac{q^2}{\hat{s}}, \frac{k^2}{\hat{s}} fixed,
\ee
the $Q_1 \bar{Q_2}$ production
cross section takes the form

\be
{\sigma}_{Q_1\bar{Q_2}} \left( s, {m_i}^2\right) = \int d^2 \q d \bar{y}
d^2 \k d
z \FF_V \left( \bar{y}, \q \right) \FF_g \left( z, \k \right)\hat{\sigma}_{Vg}
\left(\frac{\q}{\sqrt{\hat{s}}},\frac{\k}{\sqrt{\hat{s}}},
\frac{m_i^2}{\hat{s}}\right).
\ee
Here $\FF_V \left(\FF_g\right)$ are the unintegrated $V(g)$ structure
functions of
the initial particles, and $\hat{\sigma}$ is defined by a projection with
eikonal
vertices ( appropriate in the high energy limit ) of the off-shell $Vg\ra
Q_1\bar{Q_2}$ absorptive part $A^{Vg}$, in a physical gauge $A\cdot n = 0$, as
follows
\be
\ssh_{Vg}=  \frac{1}{2 s} \frac{\bar{y} z}{q^2 k^2} {\bar{p}}^{\mu}
{\bar{p}}^{\nu} p^{\rho} p^{\ss} A^{Vg}_{\mu\nu,\rho\ss}
\ee

In the following, we assume $\FF_V$ to be given by the lowest order
V-emission probability off the external fermion, i.e.,

\be
{\FF}_V \left( \bar{y}, \q \right) = \frac{\gvga}{4\pi^3} \frac{|q^2|}{
\left( |q^2| + M_V^2 \right)^2} \left( 1 -\bar{y}\right)
\ee

On the other hand, the unintegrated gluon density, $\FF_g \left( z, \k\right)$
will also contain the QCD radiative corrections that we want to take into
account, and is defined in terms of the gluon-hadron absorptive part in the
gauge $\bar{p} \cdot A = 0$ in the usual way \cite{1}

\be
\FF_g \left(z, \k\right) = \int \frac{d k^2}{\left(2 \pi\right)^4}
\frac{\k^2}{z s^2} {\bar{p}}^{\mu} {\bar{p}}^{\nu} G_{\mu \nu}\left(k,~p
\right)
\ee
so that
\be
\int_{0}^{Q^2} d^2 k \FF_g \left( z, \k\right) = g^A \left(z, Q^2\right)
\ee
is the corresponding gluon density in the initial hadron A.

The $\k$-factorized expression (2.2) shows off-shell effects
due to the dependence of $\ssh$ on the transverse momenta $\q\left(\k\right)$
of the weak boson (gluon). The $\q$-dependence was investigated in Ref.\cite{4}
and amounts to calculable corrections to the effective-W approximation
in the high energy limit. The effect is not too large, and was estimated to be
of the order of 10\% for the present value of the top mass.

On the other hand, the $\k$-dependence is enhanced for $s\gg m_t^2$,
or $z\ll 1$, by large anomalous dimension effects in $\FF_g$. They come from
QCD perturbative contributions of the form
$z^{-1} g_n\left(k^2,{Q_0}^2\right) \left( \alpha_s \log{(1/z)}\right)^n$,
which are resummed by the BFKL equation \cite{12}, and are
translated in the coefficient function by the $\k$-integration in Eq.(2.2).

In order to concentrate on strong interaction effects, we shall take in
Eq.(2.2)
the "small $\q$" limit for the weak boson, so as to translate it
into a single-$\k$ factorization formula. This is done by setting, for
$\q=O(M_V^2)$,
\be
q^\mu \simeq \bar{y} \bar{p}^\mu + \q^\mu \ra \bar{y} \bar{p}^\mu \nn
\ee
and by using in Eq.(2.3) the Ward identity \cite{4, 11}
\be
\bar{y}^2 \bar{p}^\mu \bar{p}^\nu A^{Vg}_{\mu \nu, \rho \ss} \simeq
q^\mu q^\nu A^{Vg}_{\mu \nu, \rho \ss} = {M_V}^2 A^{Gg}_{\rho \ss},
\ee
which relates the longitudinal V amplitudes to the ones for
the corresponding Goldstone bosons, and yields
\be
|q^2|\ssh_{Vg} \arrowvert_{\q^2 = 0} = \frac{M_V^2 z}{2 \bar{y}
s} \frac{p^{\rho} p^{\ss}}{|\k^2| } A^{Gg}_{\rho \ss} \equiv
\ssh_{Gg} \left( \frac{|k^2|}{z \bar{y} s}, \frac{m_i^2}{|k^2|}\right)
\ee
We then notice that the $\q^2$ integration in Eq.(2.2) can be explicitly
done around $|q| \simeq M_V$ by using Eq.(2.8) and the expression (2.4)
for $\FF_V$, to obtain
\be
\ss_{Q_1\bar{Q}_2} \left( s, m_i^2\right) = \frac{\gvga}{4\pi^2}
\frac{1}{{M_V}^2} \int \frac{d \bar{y}}{\bar{y}} \frac{dz}{z} d^2 \k
\FF_g\left(z,\k\right)
\ssh_{Gg} \left(\frac{\k^2}{z\bar{y}s}, \frac{m_i^2}{\k^2}\right)
+ O\left(\frac{1}{m^2_t}\right)
\ee

We have thus exhibited the $O\left(1/M_V^2\right)$ part of the cross section
(2.2), due to the longitudinal V polarization, that will be considered
in the following, and is also dominant in the region
$|q^2| = O\left(M_V^2\right)$, provided $M_V^2 \ll m_t^2$. The
additional $O\left(1/m_t^2\right)$ terms can in principle be fully
evalueted from (2.2) on the basis of the perturbative analysis of the full
double-$k_\perp$ cross section (2.3), but are known not to
provide large off-shell effects \cite{4} and will no longer be considered in
the following.

The single-$\k$ factorized formula (2.10) can be further elaborated to
provide a resummation formula for the corresponding coefficient function.

Firstly we define the Mellin transform in the energy variable
\be
\ss_N^{Q_1\bar{Q}_2}\left(m_i^2\right) \equiv \int_{0}^{\infty}
\frac{ds}{s} \left [\frac{s}{\left(m_1+m_2\right)^2}\right ]^{-N}
\ssh_{Q_1\bar{Q}_2}\left(s, m_i^2\right),
\ee
to rewrite Eq.(2.10) in the form
\be
\ss_N^{Q_1\bar{Q}_2}\left({m_i}^2\right)  = \frac{\gvga}{4\pi^2}
\frac{1}{N-1}
\frac{1}{M_V^2} \int d^2 k
\FF_N\left(k\right){\ssh_N}^{Gg}\left(\frac{\k^2}{M^2}, \frac{m_1}{m_2}\right),
\ee
where the $1/\left(N-1\right)$ factor comes from the $\bar{y}$ integration,
and we have defined $M = m_1 +m_2$.

Secondly, we express the $\k$-dependence of $\FF_N$ in terms of the BFKL
anomalous dimension \cite{10}
\be
\gamma_N\left(\alpha_s\right) =\gamma_N\left(\frac{\bar{\alpha_s}}{N-1}
\right) =\frac{\bar{\alpha_s}}{N-1} + 2 \zeta\left(3\right)
\left(\frac{\bar{\alpha_s}}{N-1}\right)^4 + \cdots,\left(\bar{\alpha_s}
= \frac{3 \alpha_s}{\pi}\right),
\ee
as follows
\be
\FF_N^A \left(\k \right) = \frac{1}{\pi \k^2} \gamma_N
\left(\alpha_s \right)\left(\frac{\k^2}{\mu^2} \right)^{\gamma_N
\left(\alpha_s \right)} g_{N}^{A}\left(\mu^2 \right),
\ee
where $\mu = O\left(M^2\right)$ is the factorizaton scale.

Finally, inserting Eq.(2.14) into Eq.(2.12) allows performing the
$\k$-integration in terms of the calculable $\k^2$-moments
\be
h_{N}^{Q_1\bar{Q}_2} \left(\gamma\right) = \gamma \int
\frac{d \k^2}{\k^2}\left(\frac{\k^2}{M^2}\right)^\gamma
\ssh_{N}^{Gg}\left(\frac{\k^2}{M^2}, \frac{m_1}{m_2}\right),\ee
and provides the final result
\be
\ssh_{N}^{Q_1\bar{Q}_2}\left(m_{i}^2\right)=\frac{1}{M_{V}^2}
\left(\frac{\gvga}{4\pi}\right)\frac{1}{\pi\left(N-1\right)}
h_{N}^{Gg} \left(\gamma_N\left(\alpha_s\left(M^2\right)\right)\right)
\cdot g_{N}^{A}\left(M^2\right).
\ee
Here the factor in front of $g_{N}^{A}$ provides the QCD
coefficient function for the given process. Since $\gamma_N$ in Eq(2.13)
is a known function of the effective coupling $\alpha_s/\left(N-1\right)$,
the expression (2.15), evalueted
at $\gamma_N\left(\alpha_s\left(M^2\right)\right)$, provides
the resummation formula we are looking for, once the lowest order
expression for the off-shell cross section $\ssh^{Gg}$ is given.

Note that, in deriving Eq.(2.16), we have kept, for symplicity, $\alpha_s$
frozen at its value $\alpha_s\left(M^2\right)$, and we have also
used the expression (2.14) even for $\k^2 < Q_{0}^2$, where $Q_{0}$ is a
scale defining the boundary of the perturbation approach
$\left(\alpha_s\left(Q_{0}^2\right)\leq 1\right)$.
It can be proven \cite{1,13} however, that using a RG improved expression
(2.14) and/or a full solution of the BFKL equation \cite{12} for
$\FF_N\left(k\right)$, including higher twists and possibly running coupling
\cite{14}, does not change the final result (2.16), except for subleading
terms of relative order $\alpha_s\left(M^2\right)$, which are
not considered here.

Thus our  procedure will be to first evaluate the perturbative expression for
the off-shell $\ssh^{Vg}$ and then the corresponding $h-$function,
providing the resummed coefficient.

\resection{Evaluating the coefficient function}

The hard sub-process under study $-~~V_Lg\ra Q_1\bar{Q}_2~~-$ is a
variant of the customary DIS process and has the simplifying feature of having
only abelian-type diagrams (Fig. 2). While the corresponding matrix element
is thus rather straightforword, the analytical final state integrations are
non trivial due to the unequal mass kinematics and will be sketched  in the
following.

In order to calculate $\hat{\sigma}^{Gg}$ in Eq. (2.9) it is convenient to
introduce the off-shell transverse gluon polarizations
\be
\ep_{\scriptscriptstyle IN}^\mu =
\f{1}{\k} \left(\k^\mu+\f{\k^2}{p\cdot k}p^\mu\right),~~~~~~~~~~~~~~~
\ep_{\scriptscriptstyle OUT}^\mu = (0,\mbox{\boldmath $\ep$},0),
\ee
satisfying $\ep\cdot
p=\ep\cdot \k=0$, and the longitudinal one
\be
\ep_{\scriptscriptstyle L}^\mu(k)=\f{1}{k^2}\left(k^\mu-\f{k^2}{p\cdot k}
p^\mu\right),
\ee
satisfying $\ep_\alpha\cdot k=0$, and
\be
\ep_\alpha\cdot\ep_\beta=\eta_\alpha\delta_{\alpha\beta},~~~
\sum_\alpha \eta_\alpha \ep_\alpha^\mu\ep_\beta^\nu=g_{\mu\nu}-
\f{k^\mu k^\nu}{k^2}
\ee
with $\eta_{\scriptscriptstyle IN}=\eta_{\scriptscriptstyle OUT}=
-\eta_L=-1$. By using the Sudakov parametrization
\be
k^\mu=zp^\mu-\bar{z}\bar{p}^\mu+\k^\mu=zp^\mu+\k^\mu+O\left(\f{\k^2}{\sqrt{s}}
\right);~~~k^2=-\f{\k^2}{1-z}\simeq-\k^2
\ee
it is easy to realize that, in the high energy limit,
\be
\ep_{\scriptscriptstyle L}^\mu-\ep_{\scriptscriptstyle
IN}^\mu=\f{z}{|\k|}p^\mu.
\ee
Therefore, the eikonal projection in Eq. (2.9) becomes (with
$\nu\equiv 2q\cdot p$)
\be
\hat{\sigma}_{Gg}=\f{M_V^2z}{4q\cdot k}\f{p^\rho p^\sigma}{|\k|^2}
A^{Gg}_{\rho\sigma}=\f{M^2_V}{2z\nu}(A_{LL}+A_{IN,IN})
\ee
after noticing that $A_{L,IN}=A_{IN,L}=0$ by the usual invariant decomposition
\cite{1,4}

The absorptive parts occurring in the r.h.s of Eq. (3.6) are in turn
obtained from the amplitudes
$M(G(q)+g(k)\ra Q_1(P_1,\sigma_1)+\bar{Q}_2(P_2,\sigma_2))$
by the customary integrations over phase space and spins
\be
A_{LL}+A_{IN,IN}=\sum_{\sigma_1,\sigma_2}\int d\Phi(|M_L|^2+|M_{IN}|^2).
\ee
The squared amplitudes in Fig. 2, summed over spins, turn out to be
\bea
& &\!\!\!\f{M_V^2}{2z\nu}\sum_{\sigma_1,\sigma_2}(|M_L|^2+|M_{IN}|^2)=
\nn \\
& &\!\!\!=g_s^2(C_A^2(m_1+m_2)^2+C_V^2(m_1-m_2)^2)\left(
\f{z\nu}{(m_2^2-\th)(m_1^2-\uh)}\right)+\nn\\
& &\!\!\!+g_s^2(\cvca)\f{(m_1^2-m_2^2)^2}{z\nu}\left [
\f{2m_1^2}{(m_1^2-\uh)^2}+\f{2m_2^2}{(m_2^2-\th)^2}+\right .\nn\\
& &\left .\left(
\f{z\nu}{(m_2^2-\th)(m_1^2-\uh)}\right)\left(-2+2\f{m_1^2+m_2^2}{z\nu}\right)
\right ]+\nn\\
& &\!\!\!+g_s^2(\cvca)\f{(m_1^2-m_2^2)^2\k^2}{(z\nu)^2}\left [
4\left(\f{z\nu}{(m_2^2-\th)(m_1^2-\uh)}\right)\left(1-3\f{(m_1^2-m_2^2)^2}
{(z\nu)^2}\right)+\right .
\nn\\
& &\!\!\!
-\left(\f{1}{(m_2^2-\th)^2}+\f{1}{(m_1^2-\uh)^2}\right)\left(z\nu+
6\f{(m_1^2-m_2^2)^2}{z\nu}\right)+\nn \\
& &\left .
6(m_1^2-m_2^2)\left(\f{1}{(m_2^2-\th)^2}-\f{1}{(m_1^2-\uh)^2}\right)\right ]
\eea
where $C_V(C_A)$ denote the vector (axial) $V\ra Q_1\bar{Q}_2$ coupling
constants, and we have defined $\nu=2q\cdot p$ and the Mandelstam variables
of the hard subprocess
\be
\sh=(k+q)^2,~~~\th=(q-P_1)^2=(k-P_2)^2,~~~\uh=(q-P_2)^2=(k-P_1)^2.
\ee

In the particular case of $Z_0g\ra t\bar{t}$, the heavy quark masses are
equal $(m_1=m_2=m)$ and all terms in Eq. (3.8) drop out but the first,
which yields
\be
\f{1}{2z\nu}\overline{|M_Z^2|}=\f{g_s^2}{M^2_V}4C_A^2m^2
\Big (\f{1}{m^2-\th}+\f{1}{m^2-\uh}\Big ).
\ee
Therefore, there is no explicit $\k-$dependence of $\overline{M^2}$ in this
simple case.

On the other hand, in the case of $Wg\ra t\bar{b}$, which has larger cross
section for top production \cite{15}, all the terms in Eq. (3.8) contribute
(with $m_1=m_t$ and $m_2=m_b$, say) and the $\k-$dependence is more involved.

In order to deal with the general mass configuration, it is convenient to
write the two body phase space in terms of rescaled invariants, as follows
(Appendix A)
\bea
d\Phi(1,2)&=&\f{d^4\Delta}{(2\pi)^2}\delta^+((q+\Delta)^2-m_1^2)
\delta^+((k-\Delta)^2-m_2^2)=\nn\\
&=&\f{d\tau}{8\pi}\Theta(\tau)\Theta(1-\tau)\Theta
\Big (z\nu-\k^2-\f{m_1^2}{\tau}-\f{m_2^2}{1-\tau}\Big )
\eea
where we have performed a trivial azimuthal integration, and we have defined
\be
\tau=\f{m_1^2-\th}{z\nu},~~~~1-\tau=\f{m_2^2-\uh}{z\nu}.
\ee

It is also helpful to introduce the energy type variable
\bea
& &\sigma=\tau(1-\tau)z\nu-m_1^2(1-\tau)-m^2_2\tau\geq\f{1}{4}\xi \k^2,\nn\\
& &\xi=4\tau(1-\tau)
\eea
in terms of which the phase-space boundary takes a mass-independent form:
\be
d\Phi(1,2)=\f{d\xi}{16\pi\sqrt{1-\xi}}\Theta(\xi)\Theta(1-\xi)
\Theta(4\sigma-\xi\k^2).
\ee

It turns out that, by eliminating $z\nu$, $\th$ and $\uh$ in favour of
$\sigma$ and $\xi$, Eq. (3.8) takes, after some algebra, a particularly
simple form
\bea
\f{M_V^2}{2z\nu}\overline{M^2}=& &\!\!\!\f{z\nu\xi}{4}g_s^2\left[
(C_A^2(m_1+m_2)^2+C_V^2(m_1-m_2)^2)\f{1}{(\sigma+m_1^2)(\sigma+m_2^2)}+
\right. \nn\\
& &\left. +(\cvca)\f{\xi}{4}(m_1^2-m_2^2)^2
\f{(-2\sigma+\k^2(\f{3}{2}\xi-1))}{(\sigma+m_1^2)^2(\sigma+m_2^2)^2}\right].
\eea
This expression has the remarkable property that the $\sigma -$ dependence
essentially factorizes from the $\xi -$dependence, so that the various
integrations needed are decoupled, eventually.

For instance, the $N=1$ moment of $\ssh_{Gg}$, relevant for the high energy
limit of Eq. (2.12) becomes
\bea
& &\!\!\!\ssh_{N=1}^{Gg}\Big ( \f{\k^2}{M^2},~\f{m_1}{m_2}\Big )=\nn\\
& &\!\!\!=g_s^2\int^1_0\f{d\xi}{16\pi\sqrt{1-\xi}}\int_{\f{\xi\k^2}{4}}^\infty
d\sigma\left [ \f{C_A^2(m_1+m_2)^2+C_V^2(m_1-m_2)^2}{(\sigma+m_1^2)
(\sigma+m_2^2)}+\right .\nn \\
& &\!\!\!\left . +(\cvca)\f{\xi}{4}(m_1^2-m_2^2)^2\f{\k^2(\f{3}{2}\xi-1)-
2\sigma}{
(\sigma+m_1^2)^2(\sigma+m_2^2)^2}\right].
\eea

The corresponding $\k^2-$moment, i.e., the $h-$function in Eq. (2.15)
is easily performed by doing the (linear) $\k^2-$integration before the
$\sigma$ and $\xi-$integrations, with the result:
\bea
& &\!\!\!\!\!\!h_{N=1}^{Gg}\Big (\gamma,\f{m_2}{m_1}\Big )=\f{2\as}{4}
\Big (\f{m_1+m_2}{2}\Big )^{-2\gamma}\f{1}{\gamma}B(1/2,~1-\gamma)
B(1-\gamma,~1+\gamma)\times\nn\\
& &\!\!\!\!\!\!\times\left [(C_A^2(m_1+m_2)^2+C_V^2(m_1-m_2)^2)
\f{m_1^{2\gamma}-m_2^{2\gamma}}{m_1^2-m_2^2}\right .+\nn\\
& &\!\!\!\!\!\!+
\left . \f{\cvca}{3-2\gamma}\Big (m_1^{2\gamma}+m_2^{2\gamma}-\f{2}{1+\gamma}
\f{m_1^{2(\gamma+1)}-m_2^{2(\gamma+1)}}{m_1^2-m_2^2}\Big ) \right ].
\eea

The equal mass case, relevant for $Z_0g\ra Q\bar{Q}$, is again particularly
simple because only the first term in square brackets contributes. We obtain
the mass-independent expression
\be
h_{N=1}^{Gg}(\gamma,~1)=\f{\as T_R}{4\pi}\pi C_A^2\f{\Gamma(1-\gamma)^2
\Gamma(1+\gamma)\sqrt{\pi}}{\Gamma(3/2-\gamma)}.
\ee

Of particular interest for estimating the size of scaling violations is the
behaviour of Eq. (3.16) around $\gamma=0$ (collinear limit) and $\gamma=1/2$
(saturating value \cite{16} of the BFKL anomalous dimension). The
$\gamma=0$ limit is strongly dependent on the mass ratio $r=m_2/m_1$ and has
the form
\bea
& &h_{N=1}^{Gg}(0,r)=\nn\\
& &\f{\as T_R}{2}\left [\f{2}{3}(\cvca)\Big (
2\f{1+r^2}{1-r^2}\log{\f{1}{r}}\Big )-\f{4r}{1-r^2}(C_V^2-C_A^2)
\log{\f{1}{r}}\right ].
\eea

On the other hand, the large $\gamma$ behaviour is roughly determined by
the presence of a $\gamma=1$ double pole, with behaviour
\be
h_{N=1}^{Gg}(\gamma,~r)\stackrel{\displaystyle \simeq}{\scriptscriptstyle
\gamma\ra 1}
\f{\as T_R}{\pi}
\left [\pi\Big (C_A^2+\Big (\f{1-r}{1+r}\Big )^2 C_V^2\Big )\f{1}{(1-\gamma)^2}
\right ],
\ee
to which, in Eq. (3.17), only the first term in square brackets contributes.
The saturating value at $\gamma=1/2$ is instead given by
\be
h_{N=1}^{Gg}(1/2,r)=\f{\pi^2}{2}\as\left [
C_A^2+\Big (\f{1-r}{1+r}\Big )^2\Big (C_V^2-\f{1}{6}(\cvca)\Big )\right ],
\ee
to be compared with the rough estimate obtained by setting $\gamma=1/2$ in Eq.
(3.20).

The overall behaviour of $h^{Gg}(\gamma)$ is plotted in Fig. 3 for various
mass ratios. In the equal mass case it increases from the $\gamma=0$ limit
all the way up to the double pole at $\gamma=1$. The enhancement ratio at
$\gamma=1/2$ (the asymptotic value at the BFKL Pomeron singularity) is
\be
\left . \f{h^{Gg}(1/2)}{h^{Gg}(0)}\right |_{r=1}=
\left (\f{\pi}{2}\right )^2\simeq 2.47,
\ee
a value typical of other single$-\k$ processes, like heavy flavour
photoproduction or DIS scaling violations \cite{1,3}.

Thus we conclude that the $gZ\ra Q\bar{Q}$ fusion process has rather large
resummation effects, while being disfavoured at Born level with respect to
the $gW$ and $gg$ fusion channels for top production.

On the other hand the unequal mass $gW\ra t\bar{b}$ case is peculiar
because the $\gamma=0$ coefficient (Born cross section) is enhanced,
according to Eq. (3.19), by a large factor $\sim\log{m_t/m_b}$ due, as we
shall see in Sec. 4, to a collinear singularity in the $m_b=0$ limit.
Thus the $h-$function starts decreasing away from $\gamma=0$, has a minimum,
and then is driven up to larger values by the $\gamma=1$ double pole
(Fig. 3b).

As a consequence, the enhancement ratio
\be
\left .\f{h^{Gg}(1/2)}{h^{Gg}(0)}\right |_{r\ll 1}=
\Big (\f{\pi}{2}\Big )^2\f{5}{\log{\f{1}{r}}+1}\simeq 1.49,~~~~~~~~~~~~~~
(r\approx \f{1}{30}),
\ee
is not large, and finite energy resummation effects are not likely to be
important.

The above discussion shows that the relative importance of the $Z_0g$
vs $Wg$ channel for top production is energy dependent, the latter
being dominant at low energies, but less enhanced by resummation
effects. In any case, the most important process for top production
is expected to be gluon-gluon fusion which yields a greater cross
section at low energies and is expected to be strongly enhanced by QCD
resummation effects \cite{1}  at energies of LHC type (Cf. Sec. 4C).

\resection{Process dependence of resummation effects}

We have already noticed that resummation effects due to large small$-x$
anomalous dimensions are substantially different in the $Z_0g\ra t\bar{t}$
vs $Wg\ra t\bar{b}$ cases. Comparing with the known coefficient functions of
(heavy) quark production with transverse \cite{1} and longitudinal
\cite{3} photon sources, we find the enhancement ratios $h(1/2)/h(0)$
listed in Table 1.

We can thus roughly distinguish "normally enhanced" processes
($Z_Lg,~\gamma_{\scriptscriptstyle T} g$) from not enhanced ones
($W_Lg,~\gamma_{\scriptscriptstyle L} g$).
We would like to understand this fact on the basis of the observation
\cite{1} that collinear singularities in the internal momenta may
enhance the $\k^2-$moments of the hard probe cross section.

Since we compare large $\gamma$ with small $\gamma$ moment indices, we
should distinguish at least two cases.

{\bf A) Small $\k$ enhancement.}

This occours in some massless limit, e.g. $m_q\ra 0$ in DIS-type processes,
or $m_b\ll m_t$ in the cases examined here. The collinear singularity is
due to a (nearly) massless quark exchange of virtuality $\th$ in the region
$\k^2\ll|\th|\ll Q^2$, where $Q$ (or $M$) is the hard scale. It yields the
$\log{m_t/m_b}$ factor in Eqs. (3.19) and (3.23), and a $\log{Q^2/\k^2}$
factor in the DIS case, where it provides a $1/\gamma$ pole of the coefficient
function $h_2(\gamma)$ (Cf. Ref.[3]).
This enhancement, due to a "normal" collinear behaviour in the target jet,
is important in order to assess the magnitude of the nearly on-shell (or
$\gamma=0$) region.

{\bf B) Disordered$-\k$ enhancements}

This is the typical off-shell effect
that we are exploring in the high-energy regime, in which the $\k-$phase space
opens up away from the normal collinear region mentioned before.

We shall call "disordered $\k$" region the one in which $Q^2(M^2)\ll
|\th|\ll \k^2< z\nu$ and therefore the exchanged quark aligns to
the electroweak boson rather than to the incoming hadron.
In such region the off shell cross section $\ssh_{Hg}$ for a hard source
($V$ or $\gamma$) coupled to quarks, is again dominated by a collinear
singularity as follows
\be
\ssh_{Hg}\Big (\f{Q^2}{\k^2},~\f{Q^2}{z\nu}\Big )=g_H^2\f{Q^2}{z\nu}
P_{H\ra q\bar{q}}\Big (\f{\k^2}{z\nu}\Big )\f{\as T_R}{\pi}\log{\f
{\k^2}{Q^2}},~~~~~~~~~~~~~~~~~~~~~~(\k^2\gg Q^2).
\ee

Here, however, $\bar{\tau}\equiv\k^2/z\nu$ is the Bjorken variable of the
exchanged quark, $as$ $probed$ $by$ $the$ $hard$ $gluon$,
and $P_{H\ra q\bar{q}}$ is the corresponding splitting function.

The expression (4.1) provides the correct double-pole residue at
$\gamma=1$ of the various coefficient functions in Table 1, where also
the corresponding definition of the coupling $g_H^2$ is quoted.
In fact, the $N=1$ moment of (4.1) is
\be
\ssh_{N=1}\Big (\f{Q^2}{\k^2}\Big )=\f{\as T_R}{\pi}g_H^2A_H\f{Q^2}{\k^2}
\log{\f{\k^2}{Q^2}},~~~(\k^2\gg Q^2),
\ee
\be
A_H\equiv\int_0^1d\bar{\tau}P_{H\ra q\bar{q}}\left(\tau\right ),
\ee
and thus its contribution to the $h-$function (2.14) becomes
\be
h_{N=1}^{Hg}(\gamma)\stackrel{\displaystyle \simeq}{\scriptstyle \gamma\ra 1}
\f{\as T_R}{\pi}\f{g_H^2A_H}{(1-\gamma)^2},
\ee
in agreement with the properly normalized expressions quoted in Table 1.

Here we notice that the source dependence comes from both the coupling constant
 $g_H^2$ and $A_H$, the $N=1$ moment of the splitting function.
It is easy to realize (Appendix B) that for transverse photon coefficient
functions we have (Table 2)
\be
P_{{\scriptstyle\gamma}{\scriptscriptstyle T}\ra q\bar{q}}(\bar{\tau})=
\bar{\tau}^2+(1-\bar{\tau})^2
\ee
while for longitudinal weak bosons (Goldstone bosons) we have
\be
P_{G\ra q\bar{q}}(\bar{\tau})=1,
\ee
as also quoted in Table 2.

It is clear that the double pole approximation provides most of the enhancement
factor only in those cases in which the $\gamma=0$ region is not enhanced.
This explains the "normal" enhancement of $\gamma_{\scriptscriptstyle T} g$ and
$Z_L g$ processes and the small anhancement of $W_L g$ processes.

Particular attention is needed for the $\gamma_L g$ process,
occourring in the DIS longitudinal structure function, which is
substantially different from the corresponding case in broken gauge theories.
In fact, there is no Goldstone boson in this case, and the region (A) is
not collinear singular due to the helicity flip zero at
$q^2=0$ for Breit frame scattering of massless spin $1/2$ quarks.
However, the disordered region (B) happens to have a linear collinear
divergence, being essentially of Coulomb scattering type (Appendix B).
Its contribution to $\sigma_L$ (defined by saturating with $\ep_L$ in Eq.
(B.5)) is of type
\bea
\ssh_{\gamma_L g}&=&\f{\alpha\as T_RN_f}{\pi}\f{(Q^2)^2}{z\nu}
\int_{\bar{\tau}Q^2}^{\k^2}\f{d\th}{\th^2}4\bar{\tau}^2(1-\bar{\tau})=\nn\\
&=&\f{\alpha\as T_RN_f}{\pi}\f{Q^2}{z\nu}4\bar{\tau}
(1-\bar{\tau}),~~~~~~~~~~~~~~~~~~~~~~~~~~~~~~~~~\bar{\tau}=\f{\k^2}{z\nu},
\eea
and provides a $single$ $pole$ behaviour of the $h-$function
\be
h_{\gamma_L g}(\gamma)\stackrel{\displaystyle\simeq}
{\scriptscriptstyle\gamma\ra 1}
\f{\as T_RN_f}{\pi}\f{\pi e^2 A_L}{1-\gamma}
\ee

Once again, the collinear analysis sketched above explains the weak enhancement
in this case, even if in a slightly more involved way.

{\bf C) $s-$Channel enhancement}

Finally we should recall \cite{1}, for completeness, that non abelian
gluon-gluon fusion processes show a third collinear region, the one in
the $s-$channel, which is important to estimate the size of resummation
effects, e.g. in hadroproduction of heavy flavours. This is a phase space
region of the hard subprocess $gg\ra Q\bar{Q}$ in which $M^2\ll\sh\ll z\nu$
and $M^2\ll {\bf l}^2=(\q+\k)^2\ll\k^2\simeq\q^2$. So that $both$ $\k^2$
and $\q^2$ are large and of the same order. Therefore, the massless quark
limit is relevant, the cross section $\ssh_{gg}$ behaves as $1/\sh$, and its
first moment is dominated by the $intermediate$ gluon collinear singularity
\be
\ssh_{N=1}^{gg}\left(\f{{\bf l}^2}{M^2}\right )=
\f{A_HN_c\as^2}{{\bf l}^2}\log\f{{\bf l}^2}{M^2}
\ee
for one produced flavour. Furthermore, this time
\be
A_H=\int d\tau P_{g\ra q\bar{q}}\left(\tau\right)
\ee
is related to the gluon $\ra$ quark pair splitting function in the final
state.

It turns out \cite{1} that, at extreme energies, in which both gluon
transverse momenta $\q$ and $\k$ carry a large anomalous dimension
$\gamma_N(\as)$, the behaviour (4.9) causes a $triple$ $pole$ in the
coefficient function
\be
C^{gg}(\as)\simeq (1-2\gamma_N)^{-3}A_HN_c\as^2
\ee
which occours at precisely the asymptotic value $\gamma=1/2$ because of
the double$-k$ dependence.

Of course, this implies that the non-abelian $s-channel$ region is
asymptotically dominant, so that the $gg\ra Q\bar{Q}$ channel is more
strongly enhanced than the $\gamma g$ and $Vg$ ones.

\resection{Discussion}

There are various outcomes of the analysis presented in this paper. The first
one is that the relative importance of the $Wg$ vs the $Z_0g$ fusion
processes for top production changes with energy, when scaling violations
drive the coefficient away from the $\gamma=0$ region, collinear singular
for $Wg\ra t\bar{b}$ only. However, the $gg\ra t\bar{t}$ process is even
more enhanced by scaling violations, and is thus confirmed as the most
important at energies of LHC type.

The second point is that the enhancement factors are roughly described by the
dis\-ordered$-\k$ collinear singularities, which are process dependent, since
they involve splitting functions of the hard source (transverse or
longitudinal $\gamma$'s, $Z$'s and $W$'s in the present case).
This leads to a variety of enhancement factors, and implies that it is not
really possible to eliminate $all$ large resummation effects by a
redefinition of the gluon density.

This observation is relevant for the analysis of small $x$ scaling violations
being performed at HERA. While it is fruitful to look for a factorization
scheme -or definition of the gluon density- which incorporates some
$universal$ enhancement factors, one has to live with the fact  that
resummation effects are sizeable for  $some$ processes, and may in fact
be needed to explain the small $x$ rise of structure functions.

Thus, a $comparative$ study of various processes (e.g. $F_2$,$F_L$,
heavy flavour production) is needed at both theoretical and experimental
level.

There is a final point to be noticed. The collinear analysis performed in this
paper, and  compared to exact results, provides in fact a method which can be
generalized to gluon-gluon kernels and other processes for which exact
results are not yet available. In fact, it provides a quick way to
estimating resummation effects and their process dependence, even for non
abelian hard subprocesses.
This analysis is left to future investigations.

{\bf Acknowledgements}

It is a plasure to thank Stefano Catani for a number
of enlightening discussions and suggestions. One of us (M. C.) wishes to
thank the CERN Theory Division for hospitality, while part of this work was
done. This paper is supported in part by E. C. "Human Capital and Mobility"
contract \#ERBCHRXCT930357.

\renewcommand{\theequation}{\Alph{section}.\arabic{equation}}
\appsectio{Appendix A: Unequal mass kinematics}

For the process $G(q)+g(k)\ra Q_1 \bar{Q}_2$ we use a Sudakov parametrization
of momenta, which, in the notation of Fig. 1, reads

\be
q^\mu=\bar{y}\bar{p}^\mu+\q^\mu~~~~~~~~~(q^2=-Q^2),
\ee
\be
k^\mu=zp^\mu+\k^\mu~~~~(k^2=-\k^2),
\ee
\be
\Delta^\mu=(-q+P_1)^\mu=(k-P_2)^\mu=z\tau p^\mu-\bar{y}\bar{\tau}\pb^\mu+
\dt^\mu
\ee
where $\bar{y}(z)$ is the momentum fraction of the probe (gluon) with respect
to the incoming lepton momentum $\bar{p}$ (hadron momentum $p$).

Then the mass-shell conditions $P_i^2=m_i^2$ yield
\be
\tau(1-\bar{\tau})=\f{m_1^2+(\dt+\q)^2}{z\nu},~~~~~~~~~~~~~
(1-\tau)\bar{\tau}=\f{m^2_2+(\dt-\k)^2}{z\nu}
\ee
where $\nu=2q\cdot p=\bar{y}s$. Therefore, the expression of the invariants
in terms of longitudinal and transverse variables is easily obtained.
The $\nu-$phase space is described by
\be
z\nu=\f{m_1^2+(\dt+\q)^2}{\tau}+\f{m^2_2+(\k-\dt)^2}{1-\tau}=
\f{m_1^2+(\dt+\q)^2}{1-\bar{\tau}}+\f{m^2_2+(\k-\dt)^2}{\bar{\tau}}
\ee
The first of these expressions is important for the normal collinear
kinematics,
the second one for the disordered$-\k$ region. The remaining invariants are

\bea
\!\!\!& & \hat{t} = (q-P_1)^2 = m_1^2-z\nu\tau+\q^2+2\q\cdot\dt=
-\f{m_1^2\bar{\tau}+(\dt+\bar{\tau}\q)^2}{1-\bar{\tau}}-\q^2\bar{\tau}\\
\!\!\!& & \hat{u} = (q-P_2)^2 = m_2^2-z\nu(1-\tau)+\q^2+2\q\cdot (\k-\dt)\\
\!\!\!& & \hat{s} = z\nu-(\k+\q)^2 \geq \f{m_1^2}{\tau}+\f{m_2^2}{1-\tau}.
\eea

Eqs. (A.6) and (A.7) explain the definition (3.12) of $\tau$
(for $\q^2=0$). By replacing
 (A.4) in the two-body phase space, and by performing the $\bar{\tau}$
integration first, we obtain
\be
d\Phi(1,2)=\f{d\tau}{\tau(1-\tau)}\f{d^2\dt}{2(2\pi)^2}
\delta\Big (
\f{m_1^2}{\tau}+\f{m_2^2}{1-\tau}+\f{(\dt-\tau\k)^2}{\tau(1-\tau)}+
+\k^2-z\nu \Big )
\ee
and by then performing the $d^2\dt$ integration we arrive at the
expression (3.11) of the text.

Finally Eq. (A.6)  provides the phase space boundary
\be
|\th|=|\Delta|^2\geq\bar{\tau}\Big (\f{m_1^2}{1-\bar{\tau}}+\q^2\Big )
\ee
relevant for the disordered$-\k$ kinematics, in which, by (A.5)
$\bar{\tau}\simeq\k^2/z\nu$.

\appsection{Appendix B: Longitudinal and scalar splitting functions}

Here we want to derive the splitting functions $P_H\ra q\bar{q}$ relevant
for the collinear behaviour in the disordered$-\k$ region, because
they may involve longitudinal and scalar polarizations (possibly in a broken
gauge theory), which are not usually treated in the literature.

With the notation of Fig. 1 and referring to the splitting process
$V(q)\ra$\- $Q_1(q+\Delta)$\\ $+\bar{Q}_2(-\Delta)$,
we consider the probe momentum $q^\mu$ to be off-shell and we use a Sudakov
parametrization with massless quarks
\bea
q^\mu=\bar{p}^\mu-\f{Q^2}{\nu}p^\mu,~~~& &(q^2=-Q^2,~2p\cdot\bar{p}=\nu),\\
-P_1^\mu=(q+\Delta)^\mu=(1-\bar{\tau})\bar{p}^\mu+
\f{\dt^2}{(1-\bar{\tau})\nu}p^\mu+\dt^\mu,~~~& &((q+\Delta)^2=0),\\
P^{\prime\mu}=-\Delta^\mu=\bar{\tau}\bar{p}^\mu-\f{\dt^2}{(1-\bar{\tau})\nu}-
\dt^2,~~~& &\Big (\Delta^2=-\f{\dt^2}{1-\bar{\tau}}\Big ).
\eea
where we have reabsorbed the momentum fraction $\bar{y}$ in $\bar{p}^\mu$,
for simplicity.

Here the relevant region of "disordered$-\k$" is $m_i^2,~Q^2\ll$$\dt^2\ll\k^2$,
so that we will eventually also set $Q^2/\k^2\ra 0$.

The vector boson polarizations in the Landau gauge parallel the ones in
Eqs. (3.1)-(3.3) for the gluon and are given by
\be
\ep^\mu_{\scriptscriptstyle IN}=(0,\mbox{\boldmath $\ep$}_{\scriptscriptstyle
IN},0),~~~~~~~~~~~~~~
\ep_{\scriptscriptstyle OUT}^\mu=(0,\mbox{\boldmath$\ep$}_{\scriptscriptstyle
OUT},0)
\ee
with $\mbox{\boldmath $\ep$}_{\scriptscriptstyle OUT}\cdot\dt=0$, and by
\be
\ep_{\scriptscriptstyle L}^\mu=\f{1}{Q}\Big ( q^\mu+2\f{Q^2}{\nu}p^\mu\Big ),
{}~~~~~~~~~~~~~~~~~~~~~~(\ep_{\scriptscriptstyle L}^2=1).
\ee
Correspondingly, the quark (antiquark) spinors with helicity
$\lambda(-\lambda^\prime$) take the usual form
\be
v^{\lambda^\prime}(-\Delta)=\sqrt{\bar{\tau}E_q}
\left (\begin{array}{c}\lambda^\prime\eta^{\lambda^\prime}(-\dt)\\
\eta^{\lambda^\prime}(-\dt)\end{array}\right ),
\ee
\be
u^{\lambda}(q+\Delta)=\sqrt{\bar{\tau}E_{q+\Delta}}
\left (\begin{array}{c}\lambda\chi^{\lambda}(\dt)\\
\chi^{\lambda}(\dt)\end{array}\right ),
\ee
where the relevant bispinors will be approximated, in the collinear region
by the expressions
\be
\eta^+=
\left (\begin{array}{c}1\\-\f{\ds\theta^\prime}{\ds 2}\end{array} \right ),
{}~~~~~~~\eta^-=
\left (\begin{array}{c}-\f{\ds\theta^\prime}{\ds 2}\\1\end{array} \right ),
{}~~~~~~~\theta^\prime=-\f{|\dt|}{|\vec{q}|\bar{\tau}},
\ee
and similar ones for $\chi^\pm$ with $\theta=|\dt|/|\vec{q}(1-\bar{\tau})$.
Note that off-shell effects are only relevant at order $\theta^2$.

Since we only consider, for the various initial polarizations, the leading
collinear singularity, the splitting functions can be defined
according to the probabilistic interpretation, by summing over quark
helicities in the usual way:
\be
g_H^2P_{H\ra q\bar{q}}\Big (\bar{\tau},~\f{Q^2}{\dt^2}\Big )=
\f{\bar{\tau}(1-\bar{\tau})}{2\dt^2}\sum_{\lambda,\lambda^\prime}
|M(\ep_H\ra\lambda\lambda^\prime)|^2
\ee

\newpage

{\bf Transverse polarizations}

In this well known case there is no much difference between broken and
unbroken theories. The massless quark limit is smooth, the lowest
order matrix elements, given in Table 3
\be
M_{\lambda\lambda^\prime}=\bar{u}^\lambda(q+\Delta)/\!\!\!\ep
(C_V+C_A\gamma_5)v^{\lambda^\prime}(-\Delta)
\ee
are helicity conserving and show the well known zero in the forword direction,
due to a clash with angular momentum conservation. Correspondingly, the
definition (B.9) has a finite $Q^2=0$ limit, yielding the customary
splitting function
\be
P_{T\ra q\bar{q}}(\bar{\tau})=\bar{\tau}^2+(1-\bar{\tau})^2,~~~
g_T^2=\cvca,
\ee
as quoted in Table 2. This one corresponds to the normal logarithmic
collinear singularity.

{\bf Longitudinal and scalar polarizations}

Here we should single out the broken theory case, in which the $q^\mu$ term
in the longitudinal polarization (B.5) dominates the vector exchange
contribution, for small $Q^2$ , as explained in Sec. 2.

By picking up the $q^\mu$ contribution, restoring quark masses and using
the Ward Identity
\be
M_{\lambda\lambda^\prime}^L\simeq\f{1}{Q}\bar{u}^\lambda\left [
C_V(m_1-m_2)+C_A(m_1+m_2)\gamma_5\right ]v^{\lambda^\prime}
\ee
we end up with effective (pseudo) scalar couplings,
which are singular in the $Q\ra 0$ limit and helicity violating.

The $1/Q$ factor in (B.12) is reabsorbed in the definition of the Goldstone
boson cross section $\ssh_{Gg}$ in Eqs. (2.8) and (2.9) , while for the
(pseudo) scalar matrix element we can use the massless quark kinematics as
before, thus obtaining the (pseudo) scalar values of Table 3. Note that
there is again a zero in the forword direction, due to a clash with angular
momentum conservation, which in this case would imply conserved helicity (!).

As a consequence, in the Goldstone case the collinear singularity is again
logarithmic, with constant splitting function
\be
P_{G\ra q\bar{q}}(\bar{\tau})=1,~~~~~~~~~~~g_G^2=C_S^2+C_P^2=
C_A^2\f{(m_1+m_2)^2+C_V^2(m_1-m_2)^2}{M^2_V},
\ee
as quoted in Table 1. Note that in the definition of $\ssh^{Gg}$ in Eq.
(2.9) and of $h$ in Eq. (2.16) an overall factor of $(m_1+m_2)^2/M^2_V$
has been taken out.

The longitudinal case in the unbroken theory is peculiar also. In fact,
this time the $q^\mu$ term in (B.5) yields vanishing contribution
because of current conservation, and the $p^\mu$ term is of order $Q$, but
is helicity conserving. As a consequence, there is no clash with angular
momentum conservation and no zero in the forword direction.
By using Table 3 and Eq. (B.9) we then obtain
\be
P_{L\ra q\bar{q}}=4\bar{\tau}^2(1-\bar{\tau})\f{Q^2}{|\Delta|^2},~~~
|g_L|^2=\cvca.
\ee

As a consequence, the fact that the longitudinal process has to vanish
on-shell ($Q^2=0$), is compensated by a $linear$ collinear divergence
as stated in Eq. (4.6) of the text. Integration over it, taking into account
the phase space boundary (A.10), provides an $effective$ splitting
function $4\bar{\tau}(1-\bar{\tau})$ but no logarithm and thus, upon
$\bar{\tau}-$integration the same residue $2/3$, but a single pole, in
$h(\gamma)$.This result is connected with the well-known fact that
longitudinal polarizations yield subleading logs in the collinear analysis.

To summarize, the longitudinal polarizations yield: (a) the (scalar)
helicity flipping contribution (B.12), due to current non conservation,
which in broken gauge theories yields the dominant contribution,
the probability density being of order $(mass)^2/Q^2|\Delta|^2$;
and (b) the helicity conserving contribution (B.15), the only one contributing
in the unbroken case, whose probability density is of order $1/|\Delta|^4$.
The corresponding splitting functions are given in Table 2.

\newpage
\begin{center}
{\bf \large Table 1}\\ \vspace*{1 cm}
\begin{tabular}{||c|c|c|c|c|c||} \hline
 $ process$ & $\pi g_H^2 $& $P(z)$ & $\ds\f{h^{pole}(\gamma)}{coupling}$ &
$\ds\f{h^{pole}(1/2)}{h(0)}$ &$\ds \f{h(1/2)}{h(0)}$  \\ \hline
$ Z g\ra Q\bar{Q} $&$ \f{\pi 4m^2 C_A}{M^2_Z}^2 $&$ 1 $&$
\ds\f{\as T_R}{\pi}  \f{1}{(1-\gamma)^2} $&$ 2.00 $&$ 2.47$  \\ \hline
$ W g \ra t\bar{b} $&$\scriptstyle\f{\pi m_t^2}{M^2_W}[(1+r)^2C_A^2+
(1-r)^2C_V^2] $&$
1 $&$\ds\f{\as T_R}{\pi}  \f{1}{(1-\gamma)^2} $&$ 1.33 $&$ 1.49 $ \\ \hline
$ \gamma_T g \ra Q\bar{Q} $&$ \pi e^2 $&$ z^2+(1-z)^2 $&$
\ds\f{\as T_R}{\pi} \f{2}{3}  \f{1}{(1-\gamma)^2} $&$ 1.71 $&$ 2.48$ \\ \hline
$ \gamma^\ast_T \ra q\bar{q} $&$ \pi e^2 $&$ z^2+(1-z)^2 $&$
\ds\f{\as T_R N_f}{\pi} \f{2}{3}  \f{1}{(1-\gamma)^2} $&$ 4.00 $&$ 4.00 $\\
\hline
$ \gamma^\ast_L \ra q\bar{q} $&$ \pi e^2 $&$ 4z(1-z) $&$
\ds\f{\as T_R N_f}{\pi} \f{2}{3}  \f{1}{1-\gamma} $&$ 1.00 $&$ 1.45  $\\ \hline
\end{tabular}
\end{center}

\begin{center}
{\bf \large Table 2}\\ \vspace*{1 cm}
\begin{tabular}{||c|c|c||} \hline
   i & $P^i(z)$ & $|coupling~ constant|^2 $  \\ \hline
   S & $1$ & $C_S^2+C_A^2 $ \\ \hline
   L$~^1$ & $4z(1-z)$ &$ C_V^2+C_A^2 $ \\ \hline
   IN &$ (1-2z)^2$ & $C_V^2+C_A^2 $ \\ \hline
   OUT & $1$ & $C_V^2+C_A^2  $ \\ \hline
\end{tabular}
\end{center}
\noindent
$~^1$Helicity conserving part only. The helicity flipping contribution is
equal to the scalar contribution with $C_S=(m_1-m_2)/Q$ and
$C_P=(m_1+m_2)/Q$.\\
$~^2$Effective splitting function after phase space integration (see App. B).
\newpage

\begin{center}
{\bf \large Table 3}\\ \vspace*{1 cm}
\begin{tabular}{||c|c|c|c||} \hline
  $ \lambda$ & $\lambda^\prime$ & i & $\Gamma_{ab}^i$\\ \hline
  $ +$ &$ -$ & S & $\ds-(C_S-C_P)\f{\k}{ \sqrt{z(1-z)}} $\\ \hline
  $ -$ &$ +$ & S & $\ds(C_S+C_P)\f{\k}{ \sqrt{z(1-z)}}  $\\ \hline
  $ +$ &$ +$ & L & $\ds-2(C_V+C_A)Q\sqrt{z(1-z)} $ \\ \hline
  $ -$ &$ -$ & L & $\ds2(C_V-C_A)Q\sqrt{z(1-z)} $ \\ \hline
  $ +$ &$ +$ & IN &$\ds -(C_V+C_A)\f{\k}{\sqrt{z(1-z)}}(1-2z) $ \\ \hline
  $ -$ &$ -$ & IN & $\ds(C_V-C_A)\f{\k}{\sqrt{z(1-z)}}(1-2z)  $\\ \hline
  $ +$ &$ +$ & OUT & $\ds-i(C_V+C_A)\f{\k}{\sqrt{z(1-z)}}  $\\ \hline
  $ -$ &$ -$ & OUT & $\ds-i(C_V-C_A)\f{\k}{\sqrt{z(1-z)}}  $\\ \hline
\end{tabular}
\end{center}
\newpage

{\Large\bf Figure Captions:}\\

{\bf Fig. 1:} Kinematics of the $Vg$ contribution to heavy quark production.\\

{\bf Fig. 2:} The lowest order amplitudes for the hard sub-process
$gV$$\ra$$Q_1\bar{Q}_2$.\\

{\bf Fig. 3:} The function $h(\gamma) $ for (a) $m_1=m_2$  and (b)
$m_2/m_1\simeq$$0.03$, relevant for single top production.
\newpage
\input{psfig}
\begin{figure}[htb]
\vspace*{-3.8 cm}
\centerline{\psfig{figure=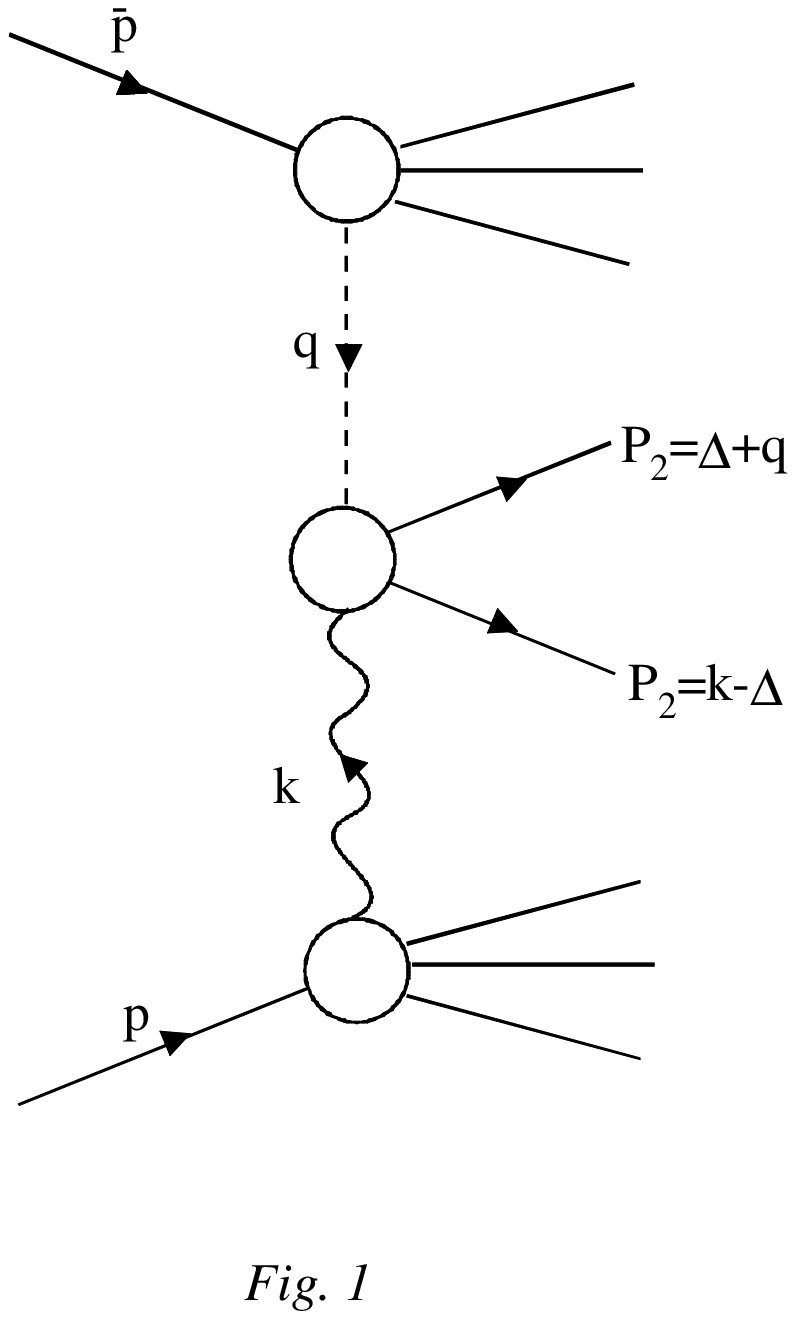}}
\vspace*{-28.5 cm}
\centerline{\psfig{figure=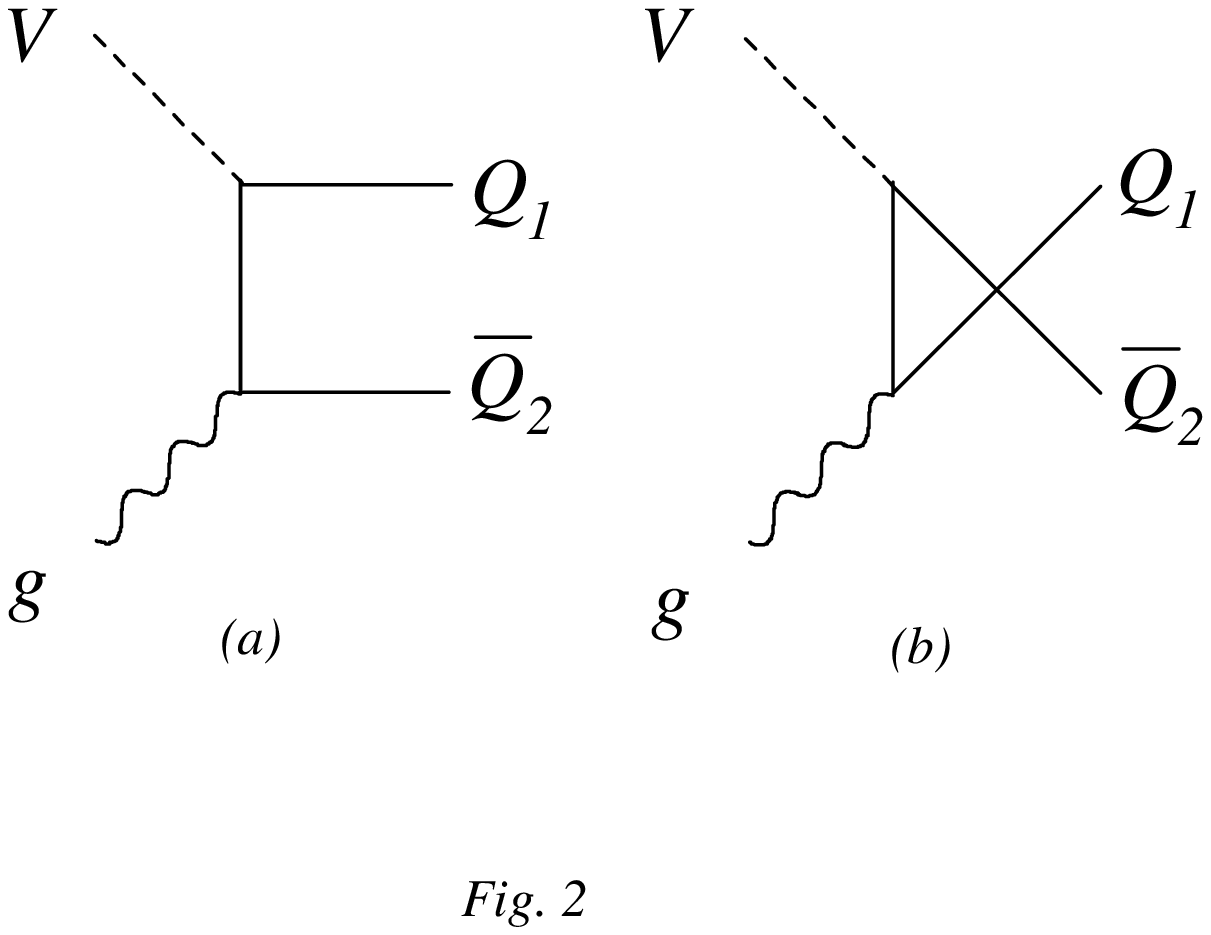}}
\end{figure}
\newpage
\input{psfig}
\begin{figure}[htb]
\centerline{\psfig{figure=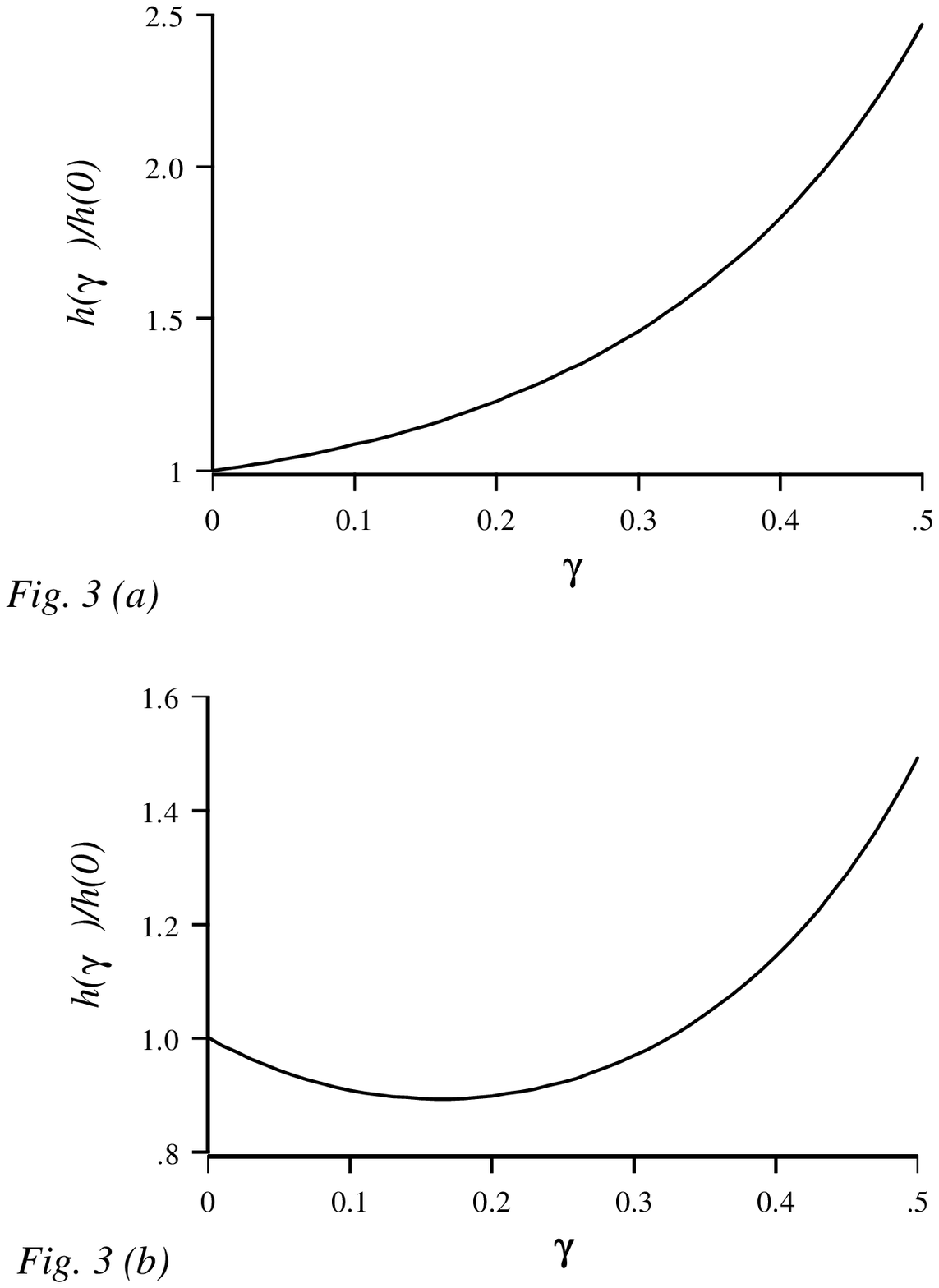}}
\end{figure}

\begin{thebibliography}{99}

\bibitem{1} S. Catani, M. Ciafaloni and F. Hautmann,
Phys. Lett. B242(1990)97; Nucl. Phys. B 366 (1991) 135.
\bibitem{2} J. C. Collins and R. K. Ellis, Nucl. Phys. B 360 (1991) 3;\\
E. M. Levin, M. G. Ryskin, Yu. M. Shable' skii and
A. G. Shuvaev, Sov. J. Nucl. Phys. 53 (1991) 657.
\bibitem{3} S. Catani and F. Hautmann, Phys. Lett. B 315 (1993) 475;
Nucl. Phys. B 427 (1994) 475.
\bibitem{4} G. Camici and M. Ciafaloni, Nucl. Phys. B 420 (1994) 615.
\bibitem{5} M. Ciafaloni, Phys. Lett. B356(1995)74.
\bibitem{6} R. K. Ellis, F. Hautmann and B. R. Webber
Phys. Lett. B 348 (1995) 582.
\bibitem{7} R. Ball and S. Forte, Phys. Lett. B351(1995)313, CERN Preprint
CERN-TH/95-184.
\bibitem{8} H1 Collaboration, T. Ahmed et al., Nucl. Phys. B 439 (1995) 471;\\
ZEUS Collaboration, M. Derrick et al., Z. Phys. C 65 (1995) 379.
\bibitem{9} R.N. Cahn and S. Dawson, Phys. Lett. B 136 (1984) 196;\\
M.S. Chanowitz and M.K. Gaillard, Phys. Lett. B 142 (1984) 85;\\
G.L. Kane, W.W. Repko and W.B. Rolnick, Phys. Lett. B 148 (1984) 367;\\
S. Dawson, Nucl. Phys. B 249 (1985) 42.
\bibitem{10} S. Cortese and R. Petronzio, Phys. Lett. B 276 (1992) 368.
\bibitem{11} A. Dobrovolskaya and V. Novikov, Z. Phys. C 52 (1991) 427;
 ITEP Preprint 95-41.
\bibitem{12} L.N. Lipatov, Sov. J. Nucl. Phys. 23 (1976) 338;\\
E.A. Kuraev, L. N. Lipatov and V. S. Fadin Sov. Phys. JETP 45 (1977) 199;\\
Ya. Balitskii and L. N. Lipatov, Sov. J. Nucl. Phys. 28 (1978) 822.
\bibitem{13} S. Catani, M. Ciafaloni and F. Hautmann, Phys. Lett.
B 307 (1993) 147.
\bibitem{14} L. V. Gribov, E. M. Levin and M. G. Ryskin,
Phys. Rep. 100 (1983) 1;\\
J. Kwiecinski, Z. Phys. C 29 (1985) 561;\\
J. C. Collins and J. Kwiecinski, Nucl. Phys. B 316 (1989) 307.
\bibitem{15}  S. S. D. Willenbrock and D. A. Dicus Phys. Rev. D 34 (1986)
155;\\
C. P. Yuan Phys. Rev. D 41 (1990) 42;\\
E. Reya et al. Proceedeings of the Large Hadron
Collider Workshop Vol. II (1990) p. 296 (CERN 90-10).
\bibitem{16} see, e. g. A. Bassetto, M. Ciafaloni and G. Marchesini,
Phys. Rep. 100 (1983) 201.
\end{thebibliography}
\end{document}